\documentstyle[preprint,aps]{revtex}
\tightenlines
\begin{document}

\draft

\title{THE E2/M1 RATIO, A STATUS REPORT}

\author{R. L. WORKMAN}
\address{Virginia Tech, Dept. of Physics, Blacksburg, U. S. A.}
\date{\today}
\maketitle

\begin{abstract}
Recent determinations of the E2/M1 ratio are reviewed. Many aspects
of these studies have been questioned. Here we focus on the selection
of a database, methods used in the extraction of this ratio,
and interpretation(s) of the results.
\end{abstract}

\section{The Data: Old vs New}

Soon after the Baryons '92 conference, the VPI group wrote a short
note\cite{VPI_old}
claiming a value of $-1.5\pm0.5$\% for the E2/M1 ratio - a
value in good agreement with the results of an independent analysis
from the RPI group\cite{DMW_91}.
Also in this paper was a value for the E2/M1
ratio obtained in a `forced fit' to preliminary $\Sigma$ data from
the LEGS group. This forced fit yielded a ratio of $-2.9$\%, a value
consistent with the most recent joint analysis of LEGS photoproduction
and Compton scattering data
($-2.85\pm0.34\pm0.21$\%)\cite{sand_trento}.

Through the 90's, the database for pion photoproduction benefited from
experimental programs at Bonn, BNL, Mainz, SAL and Yerevan. Inverse
reaction data were also measured at LANL and TRIUMF. Recent
analyses from the Mainz, LEGS, and RPI groups imply that
a much larger (larger than $-1.5$\%) E2/M1 ratio is required to fit
these more recent measurements. The Mainz group(s) find values
of about $-2.5$\%\cite{Beck,HDT}
 while the LEGS\cite{sand_trento} and RPI\cite{RPI_com}
groups have found values closer to (and slightly above) $-3$\%.

In order to understand this data dependence, several groups (LEGS,
Mainz, RPI, and VPI) have made fits using various subsets of the
existing database. The results have been very interesting. The
Mainz and LEGS groups have shown that an E2/M1 ratio close to the
VPI value can be obtained, within their fitting schemes, if particular
sets of data are used. The VPI and RPI groups have also (jointly) found
that very similar E2/M1 ratios result when identical databases
are fitted. This implies that much of the disagreement between previous
analyses has been database dependent and not model dependent.
However, one significant problem remains to be resolved.
The effect of different cross section measurements, from Bonn, Mainz and
BNL, is significant for both the E2/M1 ratio and the M1 amplitude
itself. In particular, the LEGS cross sections\cite{LEGS}
are considerably larger
than the Bonn and Mainz values. Hopefully the reason for this difference
can be understood in the near future.

\section{Extracting Amplitudes from Data}

The above mentioned agreement between different analyses holds
only in certain cases, and the following sections will mainly serve to
describe instances where results {\it do not} agree. Agreement is generally
found when a full multipole analysis is carried out. This typically allows
the variation of s- and p-wave (and possibly d-wave) multipoles over the
resonance region, with higher multipoles described by Born terms\cite{com1}.
The usually quoted E2/M1 ratio is given by Im$E_{1+}^{3/2}$/Im $M_{1+}^{3/2}$
at the energy where Re$M_{1+}^{3/2} \to 0$. The `correctness' of this
definition will be discussed in the following sections.
Here we will simply adopt
the `standard' definition as a useful point of comparison.

Watson's theorem and the dominance of a single ($M_{1+}^{3/2}$) multipole
allow stable fits over the delta resonance, apparently without the help of
double-polarization data. An issue that has been vigorously debated is
the minimal dataset and number of searched multipoles required for a
reliable extraction of the E2/M1 ratio. In Ref.\cite{Beck}, a polynomial
in $\cos (\theta )$ was fitted to the polarized-photon
$\pi^0 p$ cross sections.  The cross sections were parameterized as
\begin{equation}
{ {d\sigma}\over {d\Omega }} = {q\over k} \left( A + B\cos (\theta )
 + C\cos^2 (\theta ) \right).
\end{equation}
For polarization parallel to the beam, the coefficients are
\begin{eqnarray}
A_{||} & = & |E_{0+} |^2 + |3E_{1+} - M_{1+} + M_{1-} |^2, \nonumber \\
B_{||} & = & 2{\rm Re}\left[ E_{0+}
   \left( 3E_{1+} + M_{1+} - M_{1-} \right)^* \right], \nonumber \\
C_{||} & = & 12 {\rm Re}\left[ E_{1+} \left( M_{1+} - M_{1-} \right)^*
  \right].
\end{eqnarray}
After determining these coefficients, the E2/M1 ratio was
given by the correspondence
\begin{equation}
R_{\pi^0 p} = {1\over {12}} { {C_{||} }\over { A_{||} } } \approx R_{EM}
\end{equation}
where $R_{EM}$ is the usual E2/M1 ratio (to be evaluated at resonance)
\begin{equation}
R_{EM} = { { {\rm Re} \left( E_{1+}^{3/2} M_{1+}^{3/2*} \right) }\over
             { |M_{1+}^{3/2} |^2 } }.
\end{equation}
The simplest way to see if this is a good approximation is to just
calculate $R_{\pi^0 p}$ and compare it to the exact result, as found
from the various multipole solutions. We have done this using the RPI,
Mainz, BNL, and several VPI analyses. In some cases (for example, using
the Mainz or SM95 solutions), the difference is quite small (0.2\% or
less). However, in other cases (for example, using the RPI or BNL
solutions), the difference can be significant (about 0.3\% for BNL and
about 0.5\% for RPI). As a result, it is difficult to assign an
uncertainty to $R_{\pi^0 p}$ in advance of a full multipole analysis.

A related issue has been raised by Sandorfi\cite{sand_trento}, who has
noted that (in principle) the neglect of high partial waves can have
a significant effect on the fitted lower partial waves\cite{Donnachie}.
In the BNL analysis\cite{LEGS}, partial-waves up to $l=3$ were fitted,
with higher waves given by the Born terms. The effect of waves beyond
$l=1$ can be seen if one plots\cite{andy_priv,Beck_test}
\begin{equation}
\left( \sigma_{||} - \sigma_{\perp} \right) / \sin^2 \theta.
\end{equation}
This quantity should be angle-independent if only s- and p-waves
are important.
Deviations from this behavior are evident in the LEGS $\pi^0 p$ data
at their highest energies. This suggests that the cross sections require
non-Born contributions in the higher partial waves. In contrast, the
RPI fit, which searched only s- and p-waves - the remainder given by
Born terms, actually predicts the LEGS $\pi^0 p$ $\Sigma$ data
(a ratio of cross section measurements) at 322.6 and 333.55 MeV
with a $\chi^2$/data significantly less than unity. It should be
mentioned that similar tests, applied to the recent Mainz $\pi^o p$
measurements, showed no evidence for the importance of d-wave and
higher multipoles\cite{Beck_test}.

\section{A Plethora of Amplitudes}

The value of E2/M1 ratio determined from
Im$E_{1+}^{3/2}$/Im $M_{1+}^{3/2}$,
at the energy where Re$M_{1+}^{3/2} \to 0$, corresponds to the ratio of
amplitudes at the K-matrix pole.
A second possibility is the evaluation of an E2/M1 ratio at the T-matrix pole.
Here one forms a ratio of residues of the E2 and M1 amplitudes.
Both the individual residues and the resulting ratio are complex quantities.
Other definitions of this ratio typically require:
(a) a separation of the full
E2 and M1 amplitudes into resonance and background
pieces\cite{com2}, or (b)
a dynamical model, in which the distinction between `bare' and
`dressed' amplitudes is made. These variations will be briefly
discussed in the next section.
Just how these amplitudes are related appears to be a
model-dependent question. Since the separation of resonance and
background pieces has also been questioned, little further progress
can be made without confronting the models themselves.

\section{Model Dependence in E2 and M1}

\subsection{The K-matrix Pole}

Model dependence enters at a number of levels. Data are fitted to
phenomenological forms which satisfy the constraints of unitarity.
The precise form is not unique, and the effect of this ambiguity
on the E2/M1 ratio has been extensively studied by a number of
groups\cite{DMW_91}. Many of these recipes can be written in the
form
\begin{equation}
M = \left( M_{\rm Born} \cos \delta \; + \; M_R \right) e^{i \delta}
\end{equation}
where $M_{\rm Born}$ is the Born amplitude, $\delta$ is the corresponding
$\pi N$ phase shift, and $M_R$ is the
remainder\cite{DMW_91,Dillon,Omnes}. The first term alone
gives a qualitative description of the E2 amplitude; the second term is
dominant in the M1 amplitude.

The most important difference, between approaches falling into this class,
is the interpretation of $M_R$. If this term is associated with the
resonant part, there is no `background subtraction' at the energy where
the $\pi N$ phase shift $\to$ $\pi /2$ (since the first term vanishes).
A significantly different view has been offered in the analysis of
Aznauryan\cite{Inna}, which has also been discussed at this conference.
In this work, a particular solution of an integral equation is associated
with the background; the homogeneous solution being associated with the
resonant part. This particular solution is similar in shape to the first
term in Eq.~(6) but is larger in magnitude with an imaginary part passing
through zero at a higher energy. As a result, the remaining `resonant'
part is significantly reduced. A similar `background' curve has been
found by T.-S.H. Lee and collaborators\cite{Baryon92}. However, here the
interpretation is quite different. In Ref.\cite{Baryon92}, the
result of this background subtraction is a `bare'
photo-decay amplitude\cite{WWA}.
In Ref.\cite{Inna} the resonant part was attributed to a `dressed'
amplitude. In another recent work, the `dressed' amplitude of Lee has been
related to values extracted at the K-matrix pole\cite{Sato}.
As a result, one cannot yet claim any clear link between these
different methods.

\subsection{The T-matrix Pole}

The Mainz group has noted that, even for models and multipole fits
having very different E2/M1 ratios at the K-matrix pole, the
T-matrix pole ratio is quite stable\cite{HDT}.
This ratio of residues is determined either via the ``speed plot''
method or by continuing a model into the complex energy plane.
Results differ slightly depending on the method. However, for fits
excluding older datasets, one generally finds a
(modulus, phase) of about $(0.06,\; -125^\circ )$.

The relation between this quantity and the K-matrix pole ratio can be
understood, at least qualitatively,
if we choose to write Eq.~(6) in the form
\begin{equation}
M = \alpha ( 1 \; + \; iT_{\pi N} ) + \beta T_{\pi N}
\end{equation}
where $T_{\pi N}$ is the $\pi N$ T-matrix in a particular partial-wave,
and $\alpha$ and $\beta$ are (real) functions of energy ensuring the
correct threshold behavior. Comparing with Eq.~(6) one should recognize
that we have rewritten $\cos \delta e^{i\delta }$ in a form
$(1+iT_{\pi N})$ more convenient at the T-matrix pole. The
energy dependence of $\alpha$ (in the neighborhood of the resonance)
is essentially given by Born terms which are approximately flat for
E2 and linear for M1. The additional term proportional to $T_{\pi N}$
is required to reproduce the M1 amplitude and give a non-zero E2 at
the K-matrix pole. In fitting the phenomenological multipoles, apart
from kinematic factors, a good representation is possible with
only a constant factor for $\beta$.

Since much of the E2 energy variation is given by the first term in
Eq.~(7), it is clear that the pole residue depends largely on a
quantity which goes to zero at the K-matrix pole. Thus, the K- and
T-matrix pole ratios are sensitive to different parts of the full
multipole amplitude. In fact, the T-matrix pole ratio will remain
non-zero even in the event of a zero E2/M1 ratio at the K-matrix
pole\cite{VPI_98}.

\section{Summary}

In summary, the E2/M1 ratio evaluated at the K-matrix pole is fairly
model-independent if a consistent database is used. Inclusion of some
older Bonn neutral-pion production cross sections tends to lower the
result, while the LEGS data appear to favor a value slightly above
the Mainz result of $-2.5\%$.

The T-matrix pole ratio is quite stable but is also less obviously related
to the (generally real) model predictions. We have heard\cite{Hemmert}
that complex ratios naturally arise in ChPT calculations. This is
encouraging but at present it is not obvious that we are calculating
precisely the same quantities.

The proper way to separate background and resonance contributions
continues to generate debate. I expect this question will occupy
our attention until we better understand the connections between
quark-model and phenomenological resonance properties.

This work was supported by U.S. Department of Energy grant No.
DE-FG02-97ER41038.


\begin{thebibliography}{99}
\bibitem{com1}
The LEGS analysis varied multipoles through f-waves. These were
found to be significant even in the case of $\pi^0 p$ photoproduction,
where contributions above p-waves are usually small and attributed to
the Born terms. These higher partial-wave contributions appear to be
needed to fit the LEGS cross-section shape. The $\Sigma$ data from
LEGS can be fitted with only Born contributions beyond p-waves.
\bibitem{com2}
One should note that a resonance/background separation
is often hidden in the helicity amplitudes
$A_{1/2}= (M1+3E2)/2$ and $A_{3/2}= \sqrt{3} (M1-E2)/2$
compiled by the PDG. As a result, one should be careful
when constructing an E2/M1 ratio from these quantities.
\end{thebibliography}
\end{document}